\documentclass[preprint,3p,twocolumn,pdftex]{elsarticle}
\usepackage{lineno}
\usepackage{setspace}

\usepackage{amssymb}
\usepackage{graphicx,color}
\usepackage{float}
\usepackage{ulem}

\begin{document}

\title{Fracturing-induced fluidization of vibrated fine-powder column}
%
%
%
\author[M1,N1]{Prasad Sonar}
\author[N1]{Hiroaki Katsuragi}
 
\address[M1]{prasadrsonar@gmail.com} 
\address[N1]{Department of Earth and Space Science, Osaka University, 1-1 Machikaneyama, Toyonaka 560-0043, Japan}
\date{\today}

\begin{abstract}
We experimentally investigate the effect of vertical vibrations on the brittle behavior of fine cohesive powders consisting of glass beads of 5 $\mu \mathrm{m}$ in diameter. This is an attempt to understand a sole role of vibrations in fluidizing Geldart's group C powders, which is known for posing difficulty while fluidization. We find that the cohesive powder column can be compacted, fractured, and effectively fluidized by increasing the strengths of external vibrations. This process of vibration-induced fracturing is summarized in a full experimental phase diagram showing four distinct phases of the vibrated powder column: consolidation (CS), static fracture (SF), dynamic fracture (DF), and convective fracture (CF).
We find that the boundary separating the consolidated and fracture regimes depends on the dimensionless shaking strength, $S$.
However, in DF regime, the decompaction wave propagation speed normalized to gravitational speed is found to be independent of $S$.
In order to reach our ultimate goal of effective fluidization of group C powders, we explore geometrical parameters like container shapes, sizes, and the base conditions. We find that the circular cylinder with hemispherical base condition is the most effective container in order to achieve effective fluidization of group C powders when vibrated.
\def\keywords{\vspace{.5em}
{\bfseries\textit{Keywords}---\,\relax %
}}
\def\endkeywords{\par} 


\end{abstract}

\maketitle

\section{\label{sec:intro}Introduction}
%
%
\par Fluidization is widely used industrial process, where the granular material is transformed from solid-like state to fluid-like state \cite{kunii1991}, especially by using aeration. In chemical and pharmaceutical industries, powder fluidization is essential for efficient particle mixing and functional uniformity \cite{cano2016}.
Generally, coarse powders are easy to fluidize. But in case of fine powders, interaction forces among particles becomes larger than the gravitational force, making them cohesive and difficult to fluidize \cite{mawatari2005}. 
%
%
\par Researchers have explored and defined powders \cite{geldart1973,molerus1975,rietema1984} by studying their fluidized behavior. Some of the most common ways to fluidize powder include air or gas fluidization \cite{geldart1973,marring1994,valverde1998,cano2014,acosta2012}, using additives \cite{molerus1984,zhou2021}, and by applying vibrations to the fluidized bed \cite{marring1994,valverde2001,mawatari2015,lee2020,zhao2020,ogata2022}. Using air-fluidization approach, Geldart \textit{et al.} \cite{geldart1973} classified the powders, based on particle size and the density difference. It includes: the easily fluidized group A (exhibits dense phase expansion), B (bubbling at minimum fluidization velocity), fine and cohesive group C (very difficult to fluidize) and D (coarse grains, which forms stable spouted beds). The Geldart classification applies to dry particles fluidized with air at ambient conditions. Among those the finer and cohesive group C powders are the most difficult to fluidize. As the inter-particle forces are greater than fluid forces, such powders lift as a plug in small diameter tubes, or forms rat-hole channels, through which air/gas passes. 
To improve fluidization, it is necessary to break/fracture these stable channels. Thus, the mechanical vibrations were suggested with aeration for better fluidization of group C powders.
Recently, these suggestions were implemented and the effectiveness of vibrations was confirmed \cite{valverde2001, xu2005, mawatari2015, lehmann2019,lee2020}. It was found that group C powders indeed behave as group A powders when aerated with vibrations \cite{mawatari2015}. However, the critical vibration strength required to fracture the cohesive bulk was not discussed in detail. It is essential to find the strength requirements to fracture the consolidated bulk to understand the mechanism of fluidization in group C powders. Moreover, understanding the effect of mechanical vibrations on the physical characteristics and dynamic behavior of cohesive powders is of major industrial interest \cite{roberts1997}. 
In this paper, we experimentally examine the effect of employing pure vibrations on the effective fluidization of group C powder column, i.e. in absence of aeration. This way helps to separately understand the sudden fracturing nature of cohesive powders that potentially may aid fluidization when aerated.
%
%
\par Granular material shows large variations in behavior when subjected to external vibrations \cite{pak1994,Philippe2002,eshuis2007,zamankhan2011,kahrizsangi2016,Rosato2020}. The weak or strong vibrations can cause compaction or dilation in coarse as well as fine granular powders \cite{van1997}. On one hand, the non-cohesive coarse (group D) granular powders can be easily fluidized and its dynamic state can be governed by controlling vibrational parameters \cite{marring1994,mawatari2015}. However, on the other hand, fine cohesive powders are very difficult to fluidize as the higher inter-particle forces dominate the external forces exerted by vibrations \cite{liu2017}. 
Besides, the complex transition from consolidated powder state to fluidize powder state is arduous to interpret \cite{duran1994}. Here, we focus on investigating mechanisms of attaining effective fluidized behavior of fine cohesive powders while employing the external vibrations. While doing the systematic experiments, we also touch upon a few fascinating but unexplored phenomena particular to the cohesive powders only. This includes growth and attrition of powder agglomerates, powder slippage or adherence with the container walls, repeated bonding and debonding of fine powder grains. Each of these vibration aggravated phenomenon adds up to the complex behavior of fine powders, and needs to be studied separately. In this paper, we mention and report the occurrence of any such phenomenon that we encounter while studying effective fluidization of fine powders.
\par In past, the fluidization behavior of coarse (group D) granular material was thoroughly investigated \cite{warr1994,pak1994,wassgren1996,umbanhowar1997,lim2010,zhang2014}. In particular, Eshuis \textit{et al.} \cite{eshuis2007} employed vertical vibrations that yield a wider variety of phenomena. The mild and strong vibrational strengths were respectively characterized by non-dimensional acceleration, $\Gamma = a \omega^2 /g$, and shaking strength, $S = a^2 \omega^2 /g d$, where $a$ is vibrational amplitude, $\omega$ is angular frequency, $g$ is the gravitational acceleration, and $d$ is the particle diameter. Based on the vibration strength, the authors successfully differentiated those phenomena. Moreover, the transition between liquid-like to gas-like phase was found to be governed only by controlling $\Gamma$ and $S$. However, this study was limited to the coarse grains only. In our recently published paper, we experimentally observed that the decompaction-crack-wave velocity in dynamic phase of vibrated fine powders is governed by neither $\Gamma$ nor $S$ \cite{sonar2022}. %
In their experiments with group A powders, Pak \textit{et al.} observed convective motion that involve bubbling formation and upward motion of voids \cite{pak1994}. It was proposed that the bubbling phenomenon is related to the presence of surrounding air and occurs when $\Gamma = 6.7$. This phenomenon was later confirmed by Zamankhan \textit{et al.} \cite{zamankhan2011} and Bordbar \textit{et al.} \cite{bordbar20071} in their numerical simulations. We confirmed that this phenomenon of bubble formation is limited to dry and cohesion-less powders only \cite{sonar2022}.
Recently, when the vibrations with air fluidization approach was employed for group C powders \cite{mawatari2015}, an intermittent channel breakage and bubble formation behaviors appeared at relatively higher vibration strengths. It was found that the bed expansion ratio was low for mild vibration strength regardless of the gas velocity. The final fluidized output was the result of combined effect of aeration and vibrations. However, in presence of aeration, the influence of other parameters like shape and size of containers was ignored.
%
%
%
\par In our recently published paper \cite{sonar2022}, we experimentally studied dynamic phase of vibrated group C powders. During the phase transition, it was found that the consolidated powder fractures as the vibrational strength is increased in absence of aeration.
It was observed that the cracks in a fractured consolidate travels in an upward direction that causes an impression of the wave-propagation. Remarkably, this decompaction-wave propagation is governed by the balance between gravity and cohesion effect rather than $\Gamma$ and $S$ \cite{sonar2022}. In this paper, we continue our investigation and verify the possibility of effective fluidization of cohesive powders. To our understanding, such fluidization behavior in cohesive powders has not been studied in details. We believe, the vibration-assisted fracturing would overcome difficulties in fluidizing group C powders for industrial purpose.
%
%
%
\par The paper is arranged as follows:
In \S\ref{expt}, we introduce the experimental setup and the various parameters considered for the vibrational analysis of group C powders. Next, in \S\ref{obs}, we explain various phases we encountered when the powder column is vibrated. We summarize the powder behavior via a phase diagram. 
In \S\ref{roles}, we investigate the effect of geometric parameters like shape, size and base conditions of container on effective powder fluidization. 
\S\ref{fracture} elaborates the fracturing of cohesive powders, and its dependence on $S$.
\S\ref{disc} discusses and compares effective fluidization of cohesive powders with that of coarse powders under vibration. 
Finally, we conclude in \S\ref{con}.
%
%
%
%
\section{\label{expt}Experimental setup} 
\begin{figure*}
\centering
\includegraphics[scale=0.7]{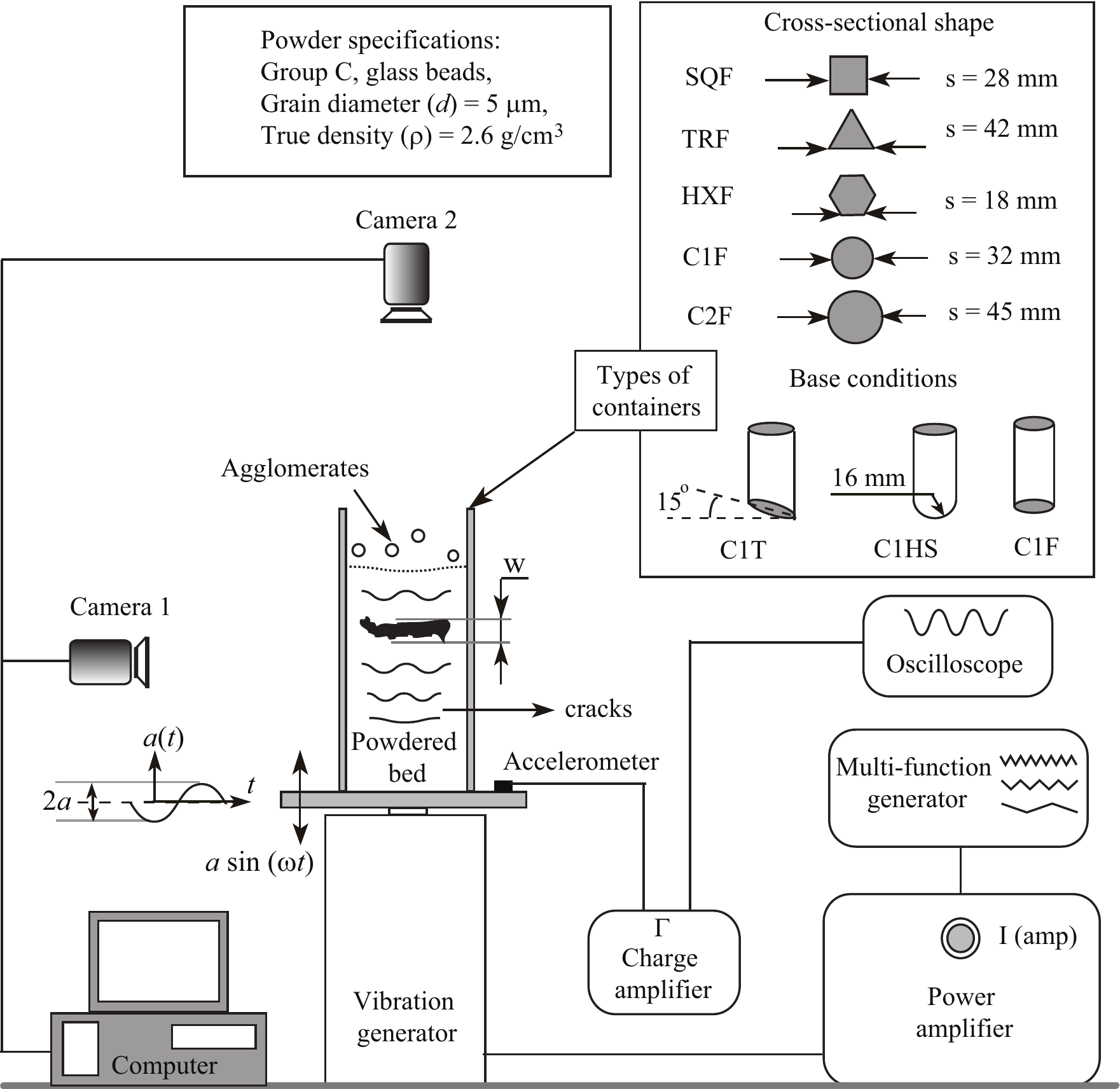}
\caption{The schematic of experimental setup, powder specifications and the types of containers used for vibrated powder column experiments. The notations used are: $a$ for amplitude, $w$ for crack width, $t$ for time, and $s$ for length of the cross-sections of container base.}
\label{fig:exptsch}
\end{figure*}
\par Figure \ref{fig:exptsch} shows the schematic of the experimental setup. The powder container is mounted on the vibration generator (EMIC 513-B/A). First,  the vibration frequency is set on the multi-function generator (NF, WF1974). This is followed by setting amplitude of vibration that is controlled by the power amplifier (EMIC 374-A). The vibration amplitude and acceleration are measured using an accelerometer (EMIC 710-C).
The glass beads of typical grain diameter, $d = 5$~$\mu$m (Potters Ballotini, EMB-10, size range is $2$--$10$~$\mu$m and true density $\rho = 2600 \, \mathrm{kg/m^3}$) are used for a model powder material. According to Geldart's chart, this material represents Group C powders \cite{geldart1973}. In such fine sized powders, the inter-particle cohesive force dominates over gravitational force~\cite{Felipe2021}.
As shown in Fig. \ref{fig:exptsch} (top right), we used various kind of powder containers having different shapes with approximately similar basal area. Moreover, we used circular cylinders of two different diameter sizes and three different base conditions.
In particular, we use circular cylinder with flat base (C1F) as our reference container, where the inner diameter and the height are $32 \, \mathrm{mm}$ and $150 \, \mathrm{mm}$, respectively. However, in some cases we also use square container with flat base (SQF) in order to see the clear planar side-view of the image. We note that throughout this paper we refer to the C1F container unless otherwise stated. This way we also confirm the observation reported by Sonar \textit{et al.} \cite{sonar2022}, where the square container is used in experiments. 
\par The experiments are conducted in following sequence. We first poured the powder into the container. The mass of powder in the container was added from $5\, \mathrm{g} \, \mathrm{to} \, 50\, \mathrm{g}$. These masses correspond to the approximate initial height of $10\, \mathrm{mm} \, \mathrm{and} \, 94\, \mathrm{mm}$ in C1F, respectively.
The initial condition for all experiments was maintained by setting the identical powder volume fraction, $\nu \approx 0.25$. The frequency of vibration was then set before performing experiments over a wide range of vibration amplitude.
Then, the powder column was vibrated in the vertical direction, i.e. the direction perpendicular to the flat base.
The ranges of vibrational parameters, amplitude and frequency $f=\omega/2\pi$ are \textcolor{black}{$0.3\, \mathrm{mm} \, \mathrm{to} \, 3.6\, \mathrm{mm}$} and $25\, \mathrm{Hz} \, \mathrm{to} \, 70\, \mathrm{Hz}$, respectively. 
We note that for every experiment, the vibrational amplitude is set to the desired value starting from the zero. Thus, the powder undergoes an identical consolidated state with increased volume fraction till it fractures. Due to the experimental limitations, the range of amplitude significantly reduces for higher frequencies. The role of frequency on powder behavior is separately discussed in our previous paper \cite{sonar2022}. Here, we mostly report results of experiments performed at $f = 30\, \mathrm{Hz}$, as it allows to incorporate wider range of vibration amplitude.
The experimental videos are recorded using front and top cameras (STC-MCCM401U3V: Equipped with a 1-inch 4-megapixel CMOS sensor) that capture images at \textcolor{black}{$89\, \mathrm{fps}$ with spatial resolution 0.058~mm/pixel and 2048 $\times$ 2048 image size}. We use fluorescent light (room light) for better image quality. 
%
%
%
\section{Observations}
\label{obs}
\par
\begin{figure*}
\centering
\includegraphics[scale=0.7]{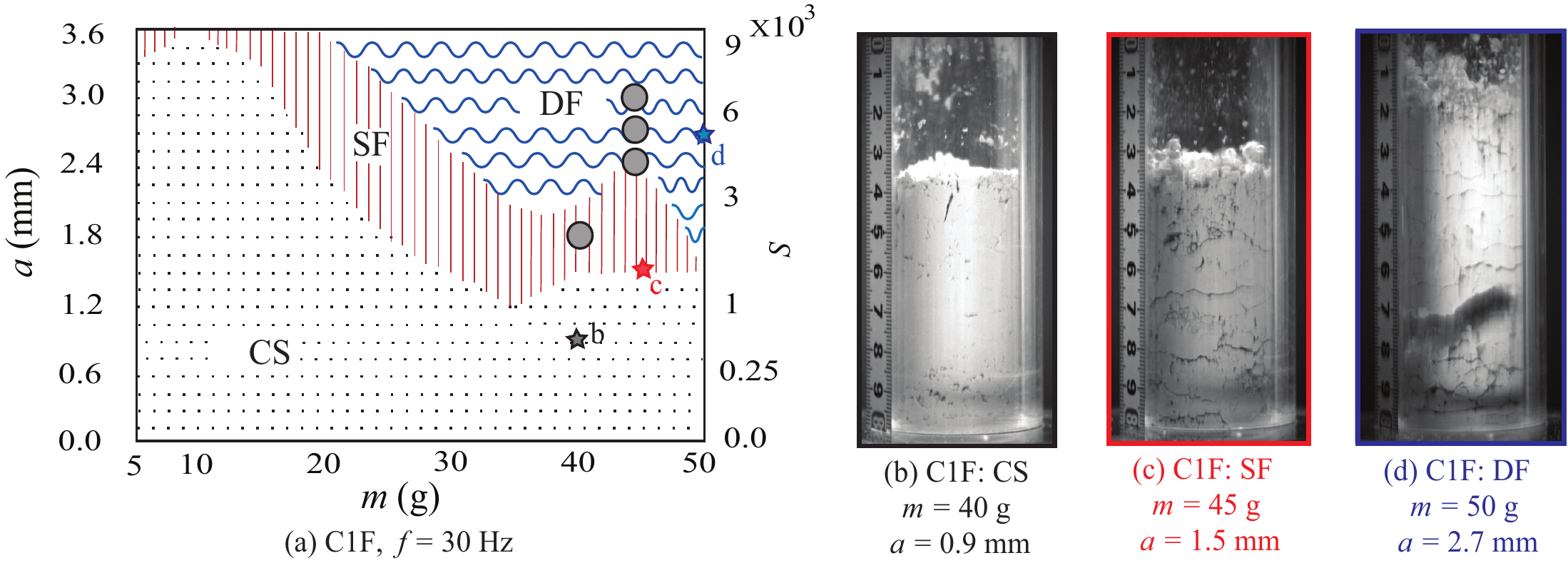}
\caption{(a) Phase diagram for circular shaped flat-based cylindrical container (C1F) with $f=30 \, \mathrm{Hz}$, RH $=40\,\%$, and the corresponding raw images of: (b) consolidated state (CS), (c) static fracture (SF), and (d) dynamic fracture (DF). The real time videos of three phases can be found in the Supplemental Material [SM] \cite{video}. The star pointers on the phase diagram indicate the data points for the corresponding raw images. The gray circular pointers represent an instant at which a large stable agglomerate was seen. The phase diagram is created based in total 120 data points that include the variation of mass, $m$, at every $5\,\mathrm{g}$ and vibration amplitude, $a$, at every $0.3\, \mathrm{mm}$.}
\label{fig:phasediagram}
\end{figure*}

 Figure \ref{fig:phasediagram}(a) shows the phase diagram of the observed phenomena when the C1F is vibrated at $f=30 \, \mathrm{Hz}$. The relative humidity, RH is $40\, \%$. We identify and distinguish three distinct phases of powder behavior: consolidation (CS), static fracturing (SF), and dynamic fracturing (DF), when various powder masses $m$ ($x$-axis) are vibrated at constant frequency and different amplitudes $a$ ($y$-axis). Figures \ref{fig:phasediagram}(b), \ref{fig:phasediagram}(c) and \ref{fig:phasediagram}(d) show the raw images of the corresponding phases CS, SF and DF on phase diagram (indicated by stars), respectively. 
These phases are described in details in our earlier paper \cite{sonar2022}. Here, we only emphasize important characteristics of the phases.
%
%
%
\par The three phases of powders are differentiated based on the deformation criteria of the void/crack. The consolidated phase is achieved when the existing voids no longer deform or completely disappear under vibrations. The ImageJ \cite{schneider2012} is used to observe deformations of voids/crack (see Fig. 8 in Sonar \textit{et al.} \cite{sonar2022}). 
As a result of weak vibrations, a single coherent bulk is obtained, where the powder column is in its most compacted and stable state. The CS regime is represented by dotted region in the phase diagram as shown in Fig. \ref{fig:phasediagram}(a). The actual image of consolidated state is shown in Fig. \ref{fig:phasediagram}(b) (see SV-CS.avi in SM for video). Due to the cohesive nature, the low mass fine powders remain consolidated even when the vibration amplitude is considerably larger than the grain size. Yet, this is not true for the case of hemispherical (C1HS) bases. We discuss this later in \S\ref{bc} while comparing effects of base conditions on powder behavior, where we show that group C powders may exhibit convection like phenomena under the influence of not only vibrations, but of different base conditions.
\begin{figure*}
\centering
\includegraphics[scale=0.6]{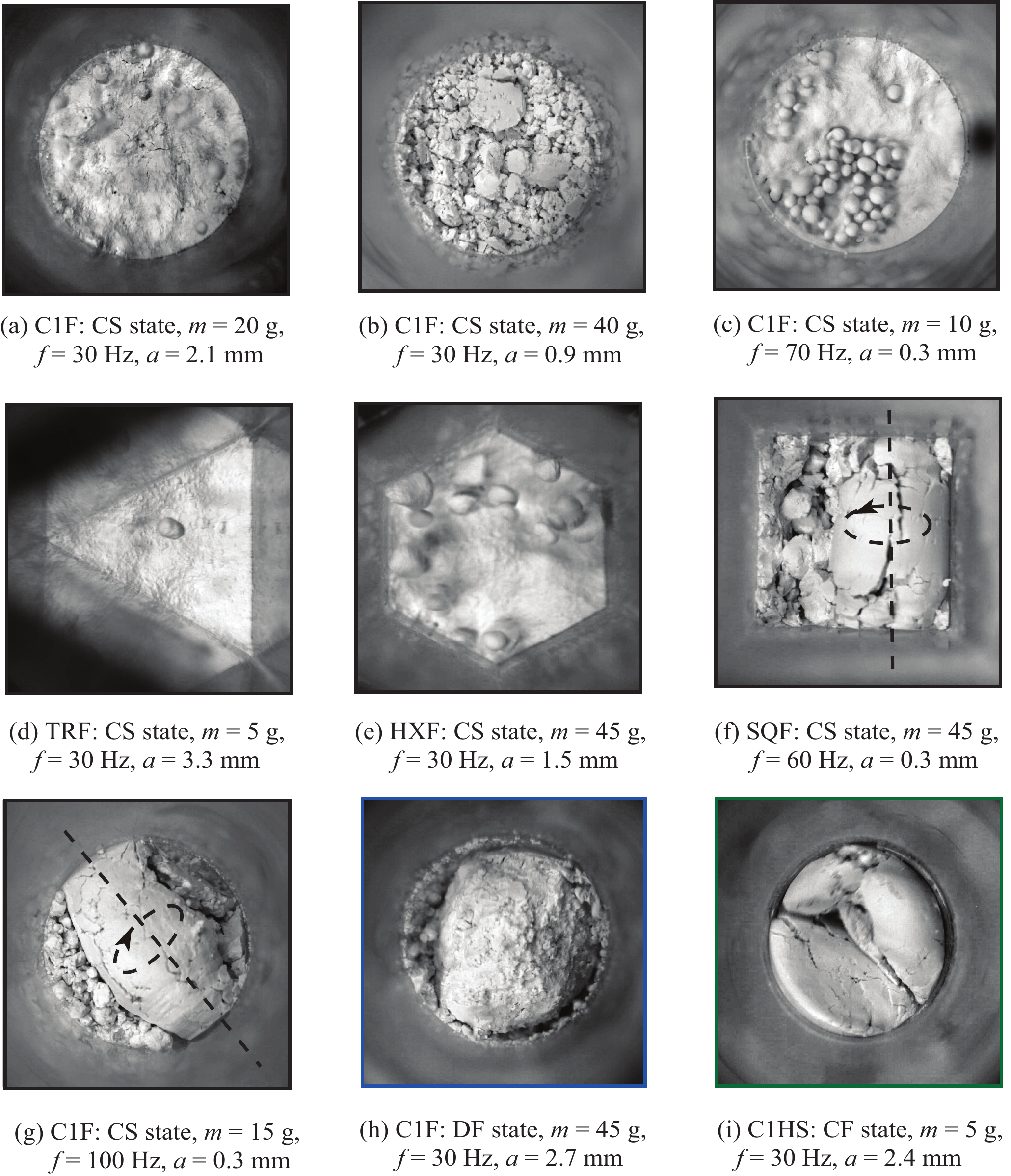}
\caption{Top-views of vibrated powder columns. CS state: (a) rough free surface, (b) fragmented free surface, (c) smooth agglomerates on the free surface, (d) and (e) smooth free surface, (f) and (g) rotating cylindrical agglomerates, DF state: (h) large agglomerate formation, and CF state: (i) convection.}
\label{fig:tvs}
\end{figure*}
\par The consolidated phases obtained as a result of weak vibrations, exhibits different kind of free surfaces. This underlines the complex behavior of cohesive granular material under vibrations. Figure \ref{fig:tvs}(a-g) shows surfaces with various textures: (a) rough, (b) fragmented, (c) small spherical agglomerated, (d, e) smooth, and (f, g) deforming cylindrical agglomerated surfaces. Typically, in cohesive powders, the loose material on the free surface forms into agglomerates. Throughout the experiments, we observed such agglomerates largely on the top free surface.
In general, small spherical agglomerates (approximately up to $8\, \mathrm{mm}$ in diameter) form at moderate vibrations (see Fig. \ref{fig:tvs}(a-e)). Nonetheless, at very low amplitudes and higher frequencies, we observed formation of cylindrical agglomerate due to convective motion of loose material on the free surface (see Fig. \ref{fig:tvs}(f, g)).
%
%
\par By further increase in vibration strength, the stable voids in consolidated coherent bulk start deforming into cracks. At a certain mass and vibration amplitude, elongated horizontal cracks start appearing in the bulk that indicate the onset of local fractures of the consolidate. This fracturing regime is named static fracture (SF) phase (shown by red colored vertical lines in Fig.~\ref{fig:phasediagram}(a)). An actual image of horizontally layered cracks generated in the bulk is shown in Fig.~\ref{fig:phasediagram}(c) (see SV-SF.avi and TV-CS-SF.avi in SM for videos). Due to this brittle fracturing nature of coherent bulk, the term \lq crack\rq\, is used instead of \lq void\rq\, as used in pre-fractured regime. For the same kind of group C powders, in fluidized bed reactor, only stable gas flow channels and the plugs are often observed when air/gas is injected. However, as observed here, the horizontal cracks formation under vibrations ensures the specific need of external vibrations for assisting fluidization, as suggested by \cite{geldart1973}. 
The CS and fractured phases are differentiated by using qualitative criteria based on deforming cracks \cite{sonar2022}. The SF state is assumed when the elongated cracks start appearing in the bulk.
\begin{figure*}
\centering
\includegraphics[scale=0.65]{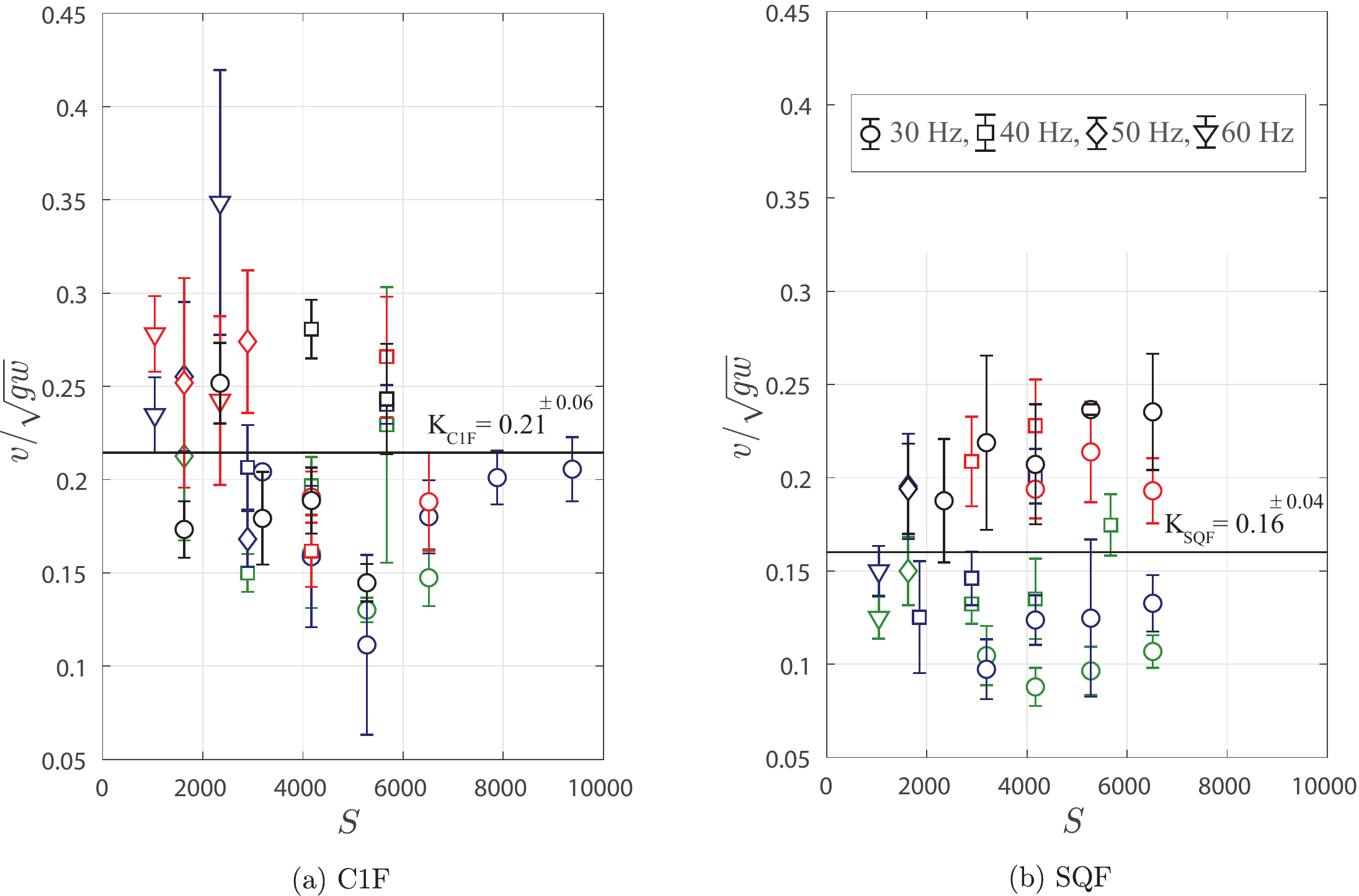}
\caption{The relation among $v/\sqrt{gw}$ and $S$ at RH $= 40 \, \%$: (a) C1F and (b) SQF. $K_{\mathrm{C1F}}$ and $K_{\mathrm{SQF}}$ respectively for (a) and (b) are the total average values of the data set that consists of various frequencies and powder masses. The powder masses $35\, \mathrm{g}$, $40\, \mathrm{g}$, $45\, \mathrm{g}$, and $50\, \mathrm{g}$ are represented by green, blue, red and black colors, respectively. Symbol shapes denote the frequency as indicated in the legend.}
\label{fig:Sind}
\end{figure*}
%
%
\par When the vibration strength is further increased, especially for increased bed mass caused a decompaction phenomenon: upward propagation of the horizontal crack~\cite{sonar2022}. The cracks in the bulk start growing wider in a successive manner from bottom to the top. This regime is called as dynamic fracture (DF) phase (shown by blue colored horizontal wavy lines in Fig. \ref{fig:phasediagram}(a) and an image in Fig. \ref{fig:phasediagram}(d), see SV-DF.avi in SM for videos). 
We refer this decompaction wave propagation in DF state as the effective fluidization of group C powders because the repeatedly traveling cracks effectively enhances fluidization by weakening the powder column.
We use quantitative criteria based on threshold area of a crack to differentiate DF phase from others phases \cite{sonar2022}. The ImageJ \cite{schneider2012} is used to track and follow the crack to find the decompaction wave velocity, $v$. The details of estimating $v$ can be found in Sonar \textit{et al.} \cite{sonar2022}. The wave velocity is then non-dimensionalized as $v/\sqrt{gw}$, where $w$ is average value of crack width, obtained as the standard deviation about the weighted mean position of the crack. 
The crack width $w$ is used as a characteristic length of the fracture, as the other important parameter, crack length, is restricted due to the container's width.
Here, we investigate DF phase characteristics using circular C1F container and confirm the finding of Sonar \textit{et al}. 
Figure ~\ref{fig:Sind} (a) and (b) shows a relation among the non-dimensionalized wave velocity $v/\sqrt{gw}$ against $S=a^2\omega^2/gd$ for circular and square cross-sections, respectively, with various frequencies, amplitudes and masses of powders at RH $= 40\,\%$. As shown in both cases, it is difficult to confirm the clear correlation as the data for $v/\sqrt{gw}$ simply scatter around a certain average value $K$. Thus, as reported in \cite{sonar2022}, we get
\begin{equation}
  \frac{v}{\sqrt{gw}}=KS^0.
  \label{eq:v_scaling}
\end{equation} 
Sonar \textit{et al.} \cite{sonar2022} concluded that in cohesive powders the decompaction wave propagation speed is solely governed by gravity, $v=K\sqrt{gw}$, where the crack width and the interval between successive cracks have the similar length scale $\sim w$. 
We confirm that the propagation speed is governed by free fall as $\sqrt{gw}$.
\par The vibration-induced fracturing causes transition from CS to DF phase, which can potentially promote fluidization. The larger $K$ value is better for efficiently fluidizing fine powder. Now we carry out parametric exploration in order to find the most useful geometrical parameters that would ensure the same. We use and compare $K$ values for various geometric parameters. Moreover, we also encounter several complex behaviors of cohesive powders in presence of vibrations. In passing, we briefly discuss these phenomena that lead us to better understand fluidization in fine cohesive powders.
%
%
%
%
\section{Parametric exploration}  
\label{roles}
\par The vibrated powder behavior depends on large number of parameters. The relative humidity is one such external parameter that can affect the dynamics of powder, especially its fluidizing behavior. However, the effect of RH was found to be negligible in our previous experiments (see Fig. 7 in Sonar \textit{et al.} \cite{sonar2022}). The experiments were performed under two different ambient conditions with relative humidity (RH) at 40\% and 85\%. The effect of RH on wave velocity and fracturing boundary was found to be negligible. Thus, for simplicity, here we assume the role of RH to be of less significance. However, throughout this paper we mention the RH values wherever required for completeness. We remind that we perform powder fluidization experiments and use only vibration without aeration. Thus, in case of aerated fluidization, the effect of RH must be studied separately.
\par The vibrational frequency is another important controlling parameter. As mentioned earlier, we performed most of the experiments at $f= 30\, \mathrm{Hz}$ due to experimental limitations. The available range of amplitude is considerably smaller for higher frequencies. As a consequence, Fig. \ref{fig:Sind} incorporates limited number of data points for higher frequencies. Besides, in our previous study, we already reported that the global structure of the phase diagrams are qualitatively similar (see Fig. 4 \cite{sonar2022}). To fracture the consolidate, it was found that the requirement of vibrational amplitude is larger for smaller frequencies. Moreover, we separately discuss and compare the brittle fracturing phenomenon in subsequent sections.
\par In this section, we investigate the effect of geometrical parameters on the wave velocity, in order to evaluate the effectiveness of powder fluidization. In past, container geometries have been studied to examine the effect of boundary conditions on convection in vibrated granular systems \cite{grossman1997,talbot2002}.
Here, we investigate the dependence of $K$ on the geometric parameters that include container shape, size and base conditions.
%
%
%
%
%
\subsection{Container shape}
\label{shape}
\begin{figure}
\centering
\includegraphics[scale=0.65]{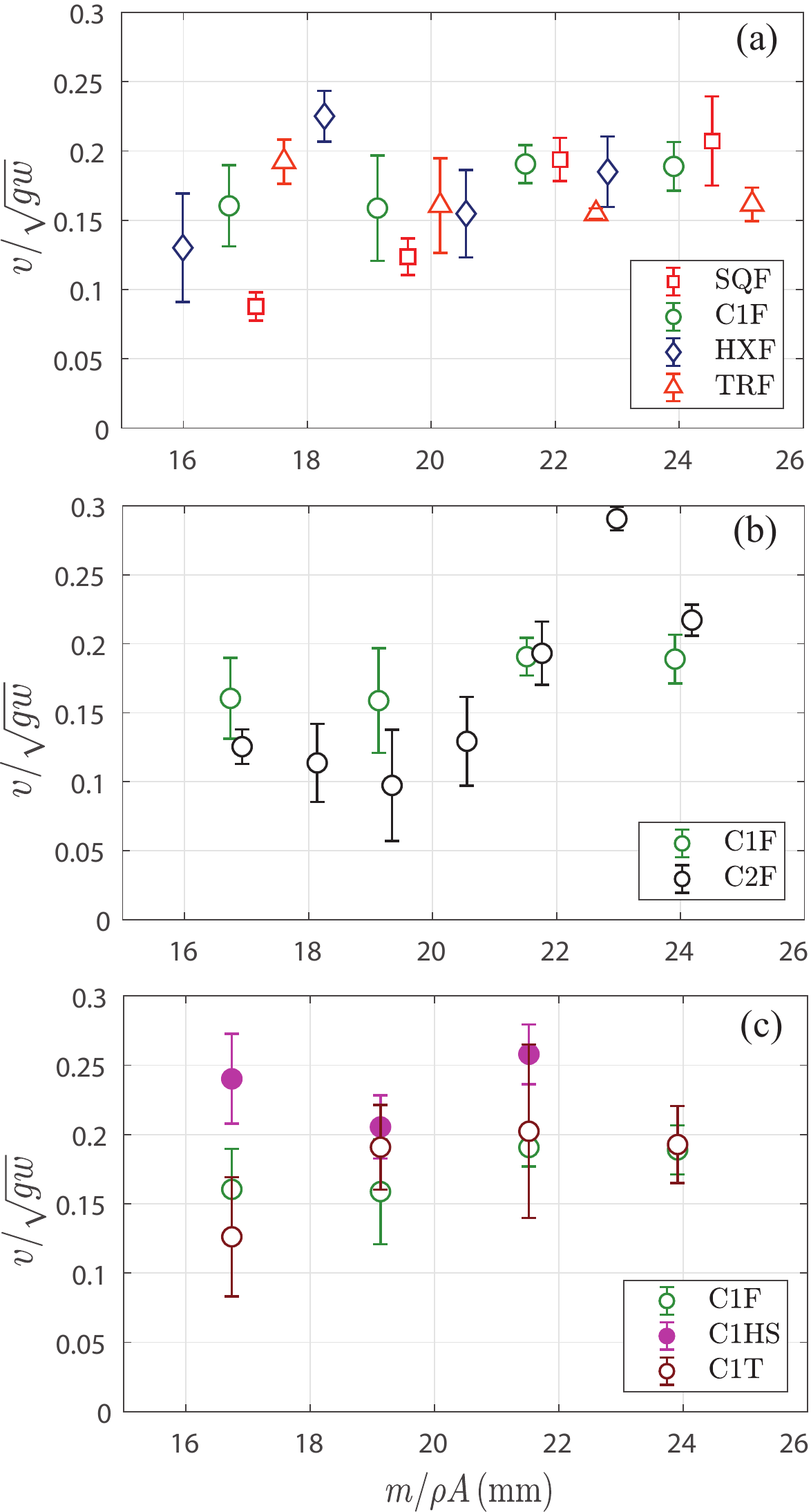}
\caption{Comparing $v/\sqrt{gw}$ for different container parameters: (a) shape, (b) size, and (c) base condition. The base is vibrated at $f = 30 \,$Hz, and at $a = 2.4 \,$ mm, i.e. $S = 4168.$}
\label{fig:vws}
\end{figure}
\par \textcolor{black}{We consider containers of different shapes with approximately equivalent basal area ($A \approx 750 \, \mathrm{mm^2}$). The shapes include circle, square, triangle and hexagon. Clearly, the shape of container affects the features of the supporting wall. For example, circular shaped container has curved walls and other containers have flat walls. However, regardless of container shape, the DF state is acquired when the bases are vibrated at $f = 30 \,$Hz, and at $a = 2.4 \,$ mm for powder mass $m>30 \,$ g. This allows us to compare $v/\sqrt{gw}$ in the DF phase. As shown in Fig. \ref{fig:vws}(a), we observed that the wave velocities among various container shapes show minor variations along the $m/\rho A$, and this variation further reduces as $m/ \rho A$ increases.}
\begin{figure*}
\centering
\includegraphics[scale=0.65]{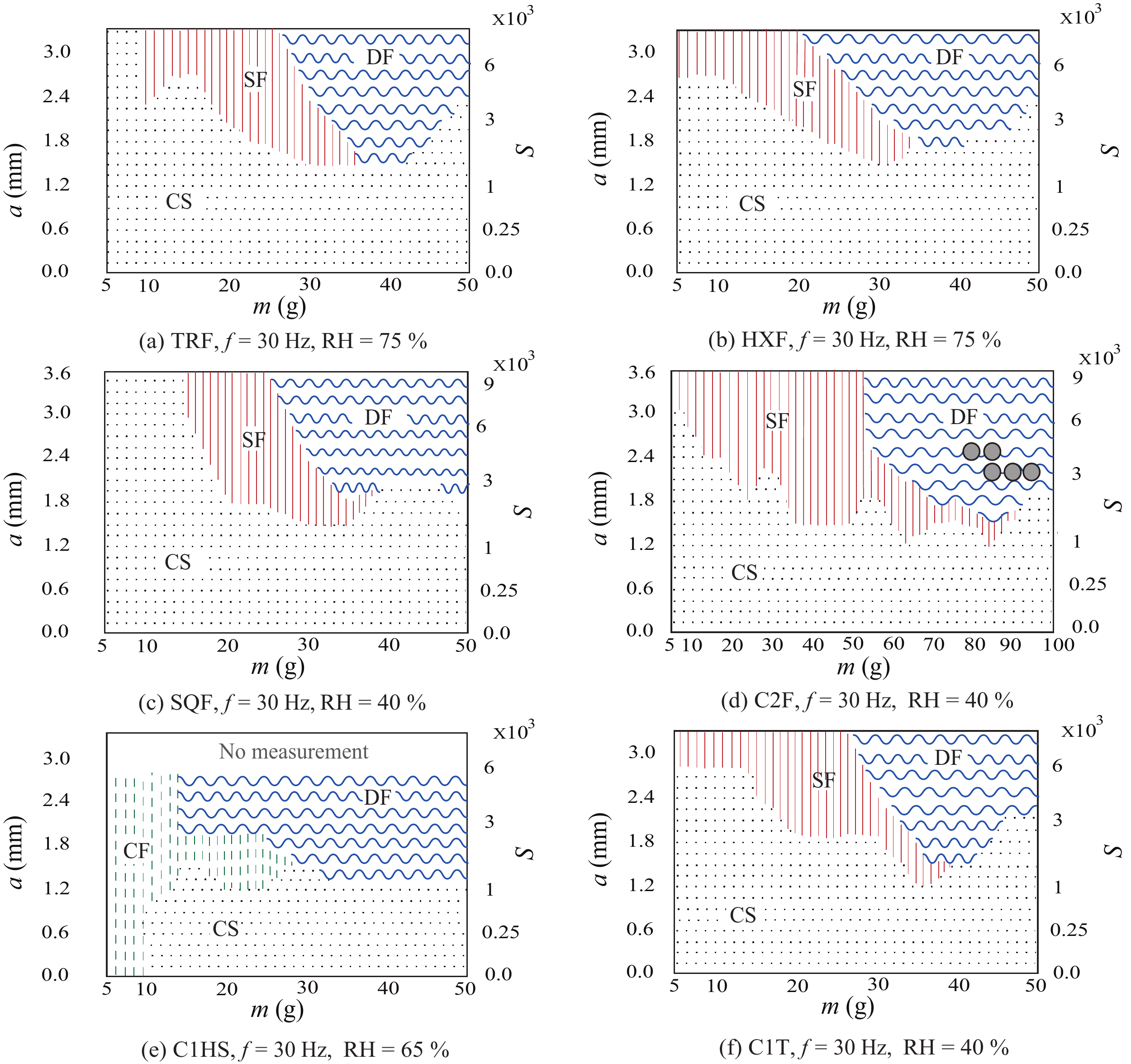}
\caption{The Phase diagrams for various types of containers vibrated at $f=30 \, \mathrm{Hz}$: (a) TRF, (b) HXF, (c) SQF, (d) C2F, (e) C1HS, and (f) C1T. The gray circular pointers represent few instances at which a large stable agglomerate was seen.}
\label{fig:phases}
\end{figure*}
\begin{figure*}
\centering
\includegraphics[scale=0.62]{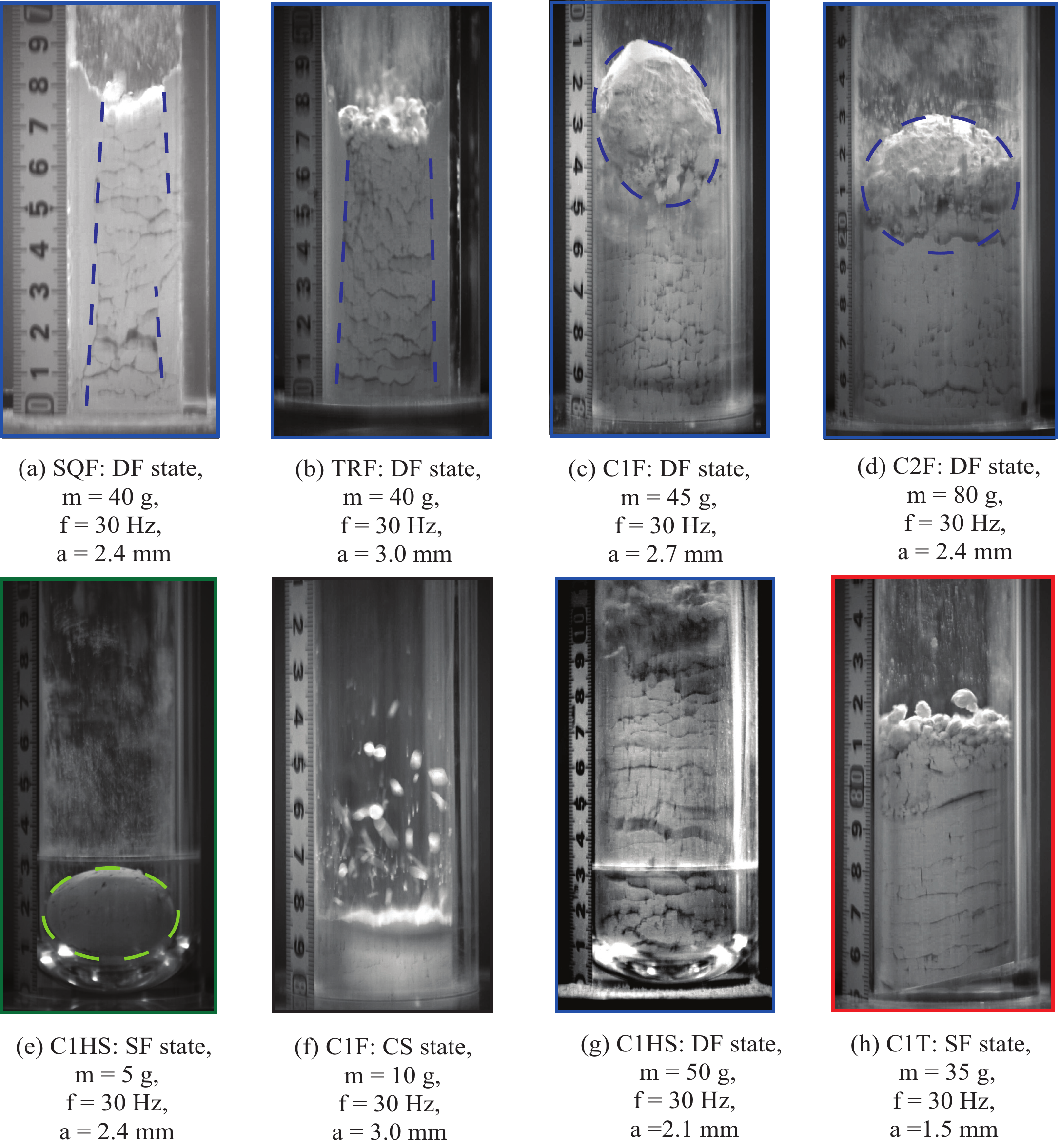}
\caption{Side-views of vibrated powder columns: Powder accumulation on (a) SQF and (b) TRF, agglomerate formation in (c) C1F and (d) C2F, low powder mass in (e) C1HS and (f) C1F, high powder mass in (g) C1HS and (h) C1T.}
\label{fig:svs}
\end{figure*}
\par \textcolor{black}{As shown in Fig. \ref{fig:phases}(a-c), the phase diagrams suggest that the transition from CS to DF phase is sudden at large $m$ except for the circular shaped container (see Fig. \ref{fig:phasediagram}(a)), where the consolidated bulk passes through SF state before reaching DF state. Thus, for higher $m$, the fracturing in circular shaped container is gradual as opposed to sudden CS to DF fracturing in flat walled containers.
We believe the container side-walls play a crucial role in the transitions from CS to fractured states. To explore this, we qualitatively compare two most commonly used shapes of the containers. 
The visual observations show that in circular container, especially for large $m$, the consolidate formed in CS state usually rotates about vertical axis when vibrational strength is increased (see CS-C1F.avi in SM for videos).
An additional amount of vibrational energy is spent on rotating the column which slides past the circular walls.
The excess motion of the powder bulk near the sidewall is sufficient to cause shearing and deformation in the bulk. Thus, initiating local cracks inside the consolidate that attains SF phase. Eventually, as the vibrational strength is further increased, crack widens and the DF phase is attained.
In contrast, in flat side-walled containers, powder accumulates at the corners, which hold the coherent bulk more rigidly. Thus, requiring more vibrational strength to initiate the crack in the consolidate (see Fig. \ref{fig:phases}(a-c)) and prolong the CS phase. However, as the vibrational strength is increased, the bulk collapses suddenly. Moreover, for flat-walled containers, powder may remain accumulated at the corners even during the DF state (see Fig. \ref{fig:svs} (a) for SQF and (b) for TRF, see TRF-acc.avi in SM for more videos). These deposits eventually break away, which otherwise causes narrow channel at the center for the propagation of decompaction wave. The adhered and accumulated powder may obstructs the upward wave motion and hence affects the wave-speed, making it less effective for fluidization compared to those obtained using circular containers (see Fig. \ref{fig:Sind}).}
In industries, adhesion or accumulation of the powder with flat surfaces are well known for creating potential challenges like clogging, or causing agglomerates and clumps. The vibrating curved walls may provide a potential solution. Thus, even though the shape factor might not be playing a significant role in DF phase, but indeed influences the phase transition that causes fracturing.
%
%
%
\subsection{Container size}
\label{size}
\par We performed identical experiments at RH $= 40\,\%$ with two different sizes of circular cylinders with inner diameters $32$ mm and $45$ mm, respectively. Figure \ref{fig:vws}(b) shows that the size of container has small influence on the wave-velocity. Figure \ref{fig:phases}(d) shows the phase diagram for the large size of circular container (C2F). For low powder masses, the fracturing occurs at low amplitudes in large container, where the side-wall area that supports powder mass is lesser compared to the vibrated basal area (bottom-wall). In contrast, in narrow containers, we reckon that the supporting side-walls play an important role of holding a cohesive bulk that needs higher vibrational strength to detach from the base. Thus, we expect the size of container to affect the fracturing especially at low mass levels.
Now, we report a drawback of employing curved walls, separately.
\subsubsection{Agglomerate formations}
\label{agg}
\par We report few instances where we observed large agglomerates, approximately more than half of container's width, formed inside both the circular cylinders, which may affect the frequency or velocity of decompaction wave. As shown in Fig. \ref{fig:svs}(c) and \ref{fig:svs}(d), the large agglomerates were seen bouncing on the free surfaces of both the narrow (see Fig. \ref{fig:tvs}(h) for the top view) and wide circular cylinders. To form such agglomerates, it requires overall higher powder mass in the container. We qualitatively observed that in circular containers, concave walls assist small or mid-sized agglomerate to grow while rotating and gathering more material. The agglomerate can grow as large as the diameter of the cylinder. Thus, here the size of container also plays a sole role in deciding the size of an agglomerate.
Besides, such large agglomerates may also affect the frequency of decompaction waves (see C1F-agg.avi in SM for more videos). The presence of large agglomerate constrains the bulk to expand and thus, halts the wave-propagation. 
The vibration of cohesive powders is expected to form agglomerates \cite{xu2005,barletta2012,kahrizsangi2016}, but its growth dependence on the curvature of container walls is not yet reported. This is one drawback of circular container that we note.
\textcolor{black}{In contrast, flat side-walled containers with square, triangular, and hexagonal shapes resist such agglomerate growth. We observed small or mid-sized agglomerate formation inside the flat walled containers. However, as the mid-sized agglomerates collide with the flat-wall, they get sheared and reduced to fragments. This process does not allow agglomerates to grow beyond a certain limit inside the flat walled containers.}
%
%
%
%
%
\subsection{Container base}
\label{bc}
\par Till now, we have worked with flat-based containers. Now, we introduce circular container with a base of hemispherical cavity (C1HS) and base tilted/inclined at $15^{\circ}$ with horizontal (C1T). Figure \ref{fig:vws}(c) compares wave-velocities for three different base conditions that appears to be almost negligible.
\par Exceptionally, we observed that the vibrated hemispherical base causes convective phenomenon in powder bulk. Among all parameters considered in this paper, the convection at the base is observed for hemispherical base containers only. Remarkably, this is true even for low mass powder columns. 
Figure \ref{fig:svs}(e) shows convection of the $5 \, \mathrm{g}$ powder inside the hemispherical base cavity when mild vibrations are employed (see Fig. \ref{fig:tvs}(i), see CF-C1HS-1.avi and CF-C1HS-2.avi in SM for videos). This is contrastive to the flat base container even at high vibration strengths, where we observed low mass powders getting consolidated (see Fig. \ref{fig:svs}(f)). However, further increase in mass over the hemispherical base causes consolidation as shown in the phase diagram (see Fig. \ref{fig:phases}(e)). The phase CF represents convection dominated fracturing phase, which is shown by green vertical dashed lines in Fig. \ref{fig:phases}(e). We note that we do not observe SF regime while using C1HS.
\par With increased vibrational strength, the consolidated bulk starts fracturing in C1HS, yet the mechanism of fracturing differs. In C1HS, the powder near the base continues to slide over the hemispherical cavity causing convective fracture near the base. Due to the curved boundaries and successive slipping of powder over walls, deformation/cracking started at the base readily reaches the top. Thus, we observe horizontal cracks sliding over the concave walls that quickens the effective powder fluidization in DF phase(see DF-C1HS.avi in SM for more video). 
Evidently, we observed fractured states with considerably high number of propagating cracks than that of with flat base. 
We qualitatively compare images of tilted, and hemispherical base containers, as shown in Fig. \ref{fig:svs}(g-h). Within C1HS, we observed many finer cracks  (see Fig. \ref{fig:svs}(g)) rather than a single wide crack as in flat based container. Thus, the fragmented state in C1HS suggests that this particular base feature enhances the chances of obtaining an improved quality of powder fluidization. 
Furthermore, unlike with C1F and C2F, we have not come across a single instant of large agglomerate formation on the top for C1HS, even though the side-walls are circular. The different failure mechanism, e.g. the presence of CF regime, underlines the greater influence of container's base condition on the effective powder fluidization.
\par In case of inclined/tilted base (C1T), we observed the behavior of powder is very similar to what was observed in C1F. Only exception being the rotation of the bulk about the vertical axis causes the cracks to tilt as shown in Fig. \ref{fig:svs}(h). Although this has almost no influence on the wave velocity in DF regime, the phase diagram suggests the change in phase boundaries (see Fig. \ref{fig:phases}(f)).
\par Finally, summarizing the role of geometric parameters, Table \ref{tab:K} compares $K$ (defined in Eq.(1) and computed by averaging ${v}/{\sqrt{gw}}$ as shown in Fig. \ref{fig:Sind}) for various experimental conditions. The highest value of $K$ is obtained for C1HS container, which promises the most effective powder fluidization.
In DF phase, the wave velocities show minor variation for various geometric parameters. However, the fractured states were observed to be influenced qualitatively by the container geometry. This include: (i) gradual and sudden phase transition in circular and flat-wall containers, respectively, (ii) fracturing in low mass powders at low strengths of vibrations using bigger sized containers, (iii) accretion and fragmentation of agglomerates in circular and flat-walled containers, respectively, and (iv) convective fracturing (CF) in hemispherical base container. This asserts that the combination of geometrical parameters may assist vibration induced effective fluidization of group C powders. Such parameters have potential to gain an industrial advantage in terms of achieving improved fluidization.
%
%
%
%
\section{Brittle fracture}
\label{fracture}
\begin{figure*}
\centering
\includegraphics[scale=0.65]{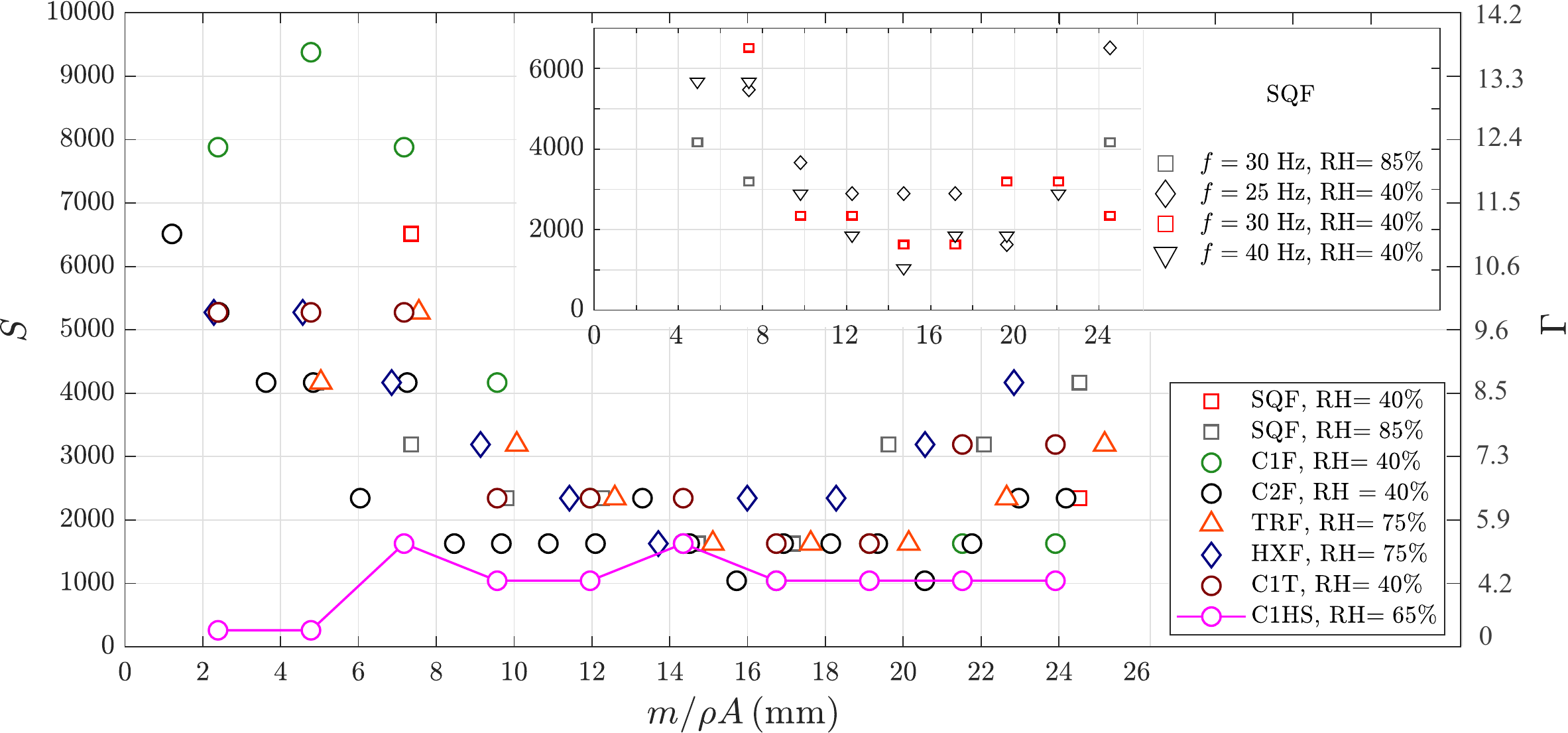}
\caption{The relation among $S$ and $m/\rho A$ at the boundary between CS and CF/SF/DF with $f = 30 \, \mathrm{Hz}$  for various geometrical parameters. Inset shows the relation among $S$ and $m/\rho A$ at for various frequencies and relative humidities for square shaped container (SQF).}
\label{fig:Sdep}
\end{figure*}
\begin{table}
\caption{\label{tab:K} Comparison of $K$ for various experimental conditions (in decreasing order).}
\begin{tabular}{cc}
\hline
\textrm{Experimental conditions} & \textrm{K} \\
\hline
	C1HS (RH = 65 \%)  & 0.23 $\pm$ 0.04   \\
	C1T (RH = 40 \%)  & 0.18 $\pm$ 0.04   \\
	C1F (RH = 40 \%)  & 0.18 $\pm$ 0.03   \\
	HXF (RH = 75 \%)  & 0.17 $\pm$ 0.04   \\
	TRF (RH = 75 \%)  & 0.17 $\pm$ 0.03   \\
	C2F (RH = 40 \%)  & 0.17 $\pm$ 0.06   \\
	SQF (RH = 40 \%)  & 0.16 $\pm$ 0.05   \\
	SQF (RH = 85 \%)  & 0.14 $\pm$ 0.03 \\
\hline
\end{tabular}
\end{table}
%
%
\par Now we analyze the fracturing boundary of the consolidated powder column. In particular, the boundary is investigated for its dependence on $S$. Figure \ref{fig:Sdep} shows the comparison of boundaries separating the consolidated and fractured phases, when the base is vibrated at $30$ Hz for all the experimental conditions and parameters we used. We vary the amplitude and plot the relation among the corresponding shaking strength $S$ and $m/\rho A$. The CS and SF/DF/CF phases lie on the lower and upper sides of the boundaries, respectively. As shown, we observe a reasonable collapse forming an \lq\lq U\rq\rq \, shaped profile for all the geometric parameters considered, except for the one with hemispherical base condition (C1HS, magenta line). We find that the minimum vibrational strength required for fracturing a cohesive bulk powder is around $S \simeq 2000$ that corresponds to $\Gamma \simeq 6$. 
\par {\color{black}In general, the shape of fracturing boundary may depend upon: 1. area of supporting wall and the base, 2. voids present inside the bulk, and 3. mass of the powder.} We observed that low mass levels solidify easily and require considerably high vibrational strength to fracture. 
This requirement of $S$ reduces with increased mass level (for $m/\rho A \, \le \, 10 $ mm).
For the fine cohesive powders used here, the tendency to adhere to the base and sidewall is high. Such low masses form a solid powder column which is supported by sidewalls and moves with the vibrated base without getting detached. For coarse granular material, the detachment condition is $\Gamma > 1$ \cite{pak1994,eshuis2007}. However, here the bonds within grains and with walls are strong enough to overcome high vibrational strength. In addition, the smaller voids still present inside the bulk are further stabilized by having supporting walls. At very high amplitudes these small voids finally open to cause deformation or localized fracture in the bulk. Due to the shallow column thickness at low masses, fracture causes local deformations or eruptions and we do not see crack formation. {\color{black}This fracturing in low mass consolidates is dominated by the adhesion between powder grains and the walls. However, in a container with larger base size (C2F), force exerted by vibrated base overcomes the adhesive forces on supporting walls to fracture the shallow coherent mass.}
\par For moderate mass level (for $10 \, \textrm{mm} \le \, m/\rho A \, \le \, 20 $ mm), the fracturing occurs at considerably low vibrational strength. This happens when forces exerted by the base is dominant over the sidewall friction/adhesion. The moderate mass accommodates higher number of voids within the bulk.
The cracking first starts from the existing voids inside the consolidate. In circular containers, powder bulk rotates inside the container while cracking, thus, causing shearing and finer cracks. In contrast, the flat walled containers keep the coherent powder bulk intact for longer time. However, vibrated base in presence of more number of voids dominates the support provided by wall to cause the fracture at considerably lower strengths. Thus, for moderate heights, the cohesive bulk powder fractures at lowest vibrational strength. The role of frictional sidewalls can be investigated in future by observing the range of mass level at which the powder bulk fractures at low vibrational strengths. Indeed, boundary condition (base shape) affects the shape of phase boundary (C1HS curve in Fig. \ref{fig:Sdep}). Further experiments with various boundary conditions are crucial future work.
\par When the mass in the container exceeds $40$ g (for $m/\rho A \, \ge \, 20 $ mm), the gravitational force dominates the induced kinetic energy. At low amplitudes, the powder collapses and forms consolidate, which is dominated by gravitational acceleration and supported by sidewalls. This requires additional vibrational strength to mainly overcome gravitational acceleration and wall resistance to cause fracture. Thus, the large mass consolidate fractures at higher vibrational strength. 
The inset in Fig. \ref{fig:Sdep} confirms that the similar \lq\lq U\rq\rq \, shaped trend can be observed for varying vibrational frequencies and RH.
\par In summary, over the number of experiments considering various parameters, we observed a reasonable collapse that suggests the $S$-dependent fracturing. The fine and cohesive nature of powder along with various geometric parameters make it difficult to achieve precise collapse. Besides, the challenges in identifying finer cracks make it harder.
The scaling with $S$ is important because it would allow us to design the vibrating system for minimum specific power/strength requirement that would fracture the consolidate. Now, we know that for moderate mass levels powder bulk fractures at $S \simeq 2000$. This information makes it easier to fluidize the group C powders when vibrations are employed with aeration.
Also, the decompaction wave propagation can then be initiated with increased $S$, and can further improve fluidization of powder with aeration \cite{sonar2022}. Estimating the effectiveness of such system is our objective for the future research.
%
%
\section{Discussion}
\label{disc}
In past, vibrated granular beds are studied by many researchers \cite{pak1994, umbanhowar1997, eshuis2007, taberlet2007}. Here, we discuss our findings and compare those with relevant literature available. 
Particularly, we compare our work with Eshuis \textit{et al.} \cite{eshuis2007}, who experimented with group D powders in quasi-two-dimensional granular system ($1$ mm grain size, $a = 2$ to $4$ mm, bed height of $15$ grain diameters). Apart from this, we qualitatively relate decompaction wave propagation with the bubbling phenomenon, which was observed in group B powders by Pak \textit{et al.} \cite{pak1994} ($0.2$ mm grains size,  $a = 4$ mm, bed height of $40$ mm).
\par Dry granular bed behaves like a solid, not detaching/separating with vibrated bed until $\Gamma \leq 1$ \cite{pak1994, eshuis2007}. To detach, the vibrated bed must have downward acceleration that overcomes gravity as well as wall friction/adhesion. The criterion used for the detachment condition \cite{duran2012} is $\Gamma = 2 -e^{-\chi}$, with $\chi = R \mu_s \zeta$, where $R$ is coefficient of redirection toward the wall, $\mu_s$ is static friction coefficient, and $\zeta$ is ratio of contact area to the cross sectional area. Once the detachment condition is fulfilled, the dry granular bed bounces like a single particle. 
However, this is not the case for fine cohesive powders that we use.
Also, the cohesive forces among the grains and walls form a very strong bond to be overcome by the corresponding vibrational strength. Thus, we observe compaction that causes increased volume fraction and results in consolidated powder bulk. We call this as CS phase. Especially on flat vibrated bases, low powder masses are easy to consolidate and remain so for high vibrational strengths.
\par When vibrational strength is further increased ($\Gamma > 1$), for various frequencies and parameters, we observed cracking in the consolidated bulk. Figure \ref{fig:Sdep} shows a data collapse suggesting the dependence of a fracturing phase boundary on $S$. The consolidate was found to be fractured at around $S \simeq 2000$. This corresponds to $\Gamma \simeq 5.9$. Since $d$ is fixed, $S$ and $\Gamma$ can be translated.
Surprisingly, for the similar $\Gamma$, Eshuis \textit{et al.} \cite{eshuis2007} observed the onset of undulation regime, where the granular bed shows standing wave patterns similar to a vibrating strings. Each collision with the vibrating base causes a shock wave through the granular bed. Remarkably, using group B powders, Pak \textit{et al.} \cite{pak1994} reported a bubbling effect, defined as upward moving voids, reminiscent of bubbles in fluidized bed, at $\Gamma = 6.7$. As the bubbles are created at the base and move upwards, their mean size grows. This was verified using numerical simulation by other researchers \cite{zamankhan2011, bordbar20071}. Sonar \textit{et al.} \cite{sonar2022} confirm the bubbling related phenomenon by conducting experiments with vibrated group A ($d = 50$~$\mu$m and true density $\rho = 2600 \, \mathrm{kg/m^3}$) powder column. 
The onset of bubbling and undulations obtained by Pak \textit{et al.} \cite{pak1994} and Eshuis \textit{et al.} \cite{eshuis2007}, respectively for coarser granular media coincides with cracking/fracturing of vibrated cohesive powders, especially for moderate masses. The mechanism of forming voids/bubbles in coarse granular media is governed by the presence of compressed air, which is trapped in the material and expands during the detachment process \cite{pak1994,zamankhan2011}.
However, in fine cohesive powders, we do not see any detachment at the base. Instead we observe horizontal crack formations throughout the bulk, which occurs as the cohesive bonds among the grains break. This we call brittle fracturing. The mechanism of fracture in cohesive powder is yet to be studied. Also, the role of trapped air inside the cohesive material is yet to be verified.
Although, the coarse and fine cohesive grain powders deform/fracture in ductile and brittle manner, respectively, the phase change occurs at moreorless similar $S$ and $\Gamma$ values.
\par Eshuis \textit{et al.} observed Leidenfrost effect when the vibration strength is further increased. During this phase, a dense particle regime is elevated and supported by dilute gaseous layer underneath. The vibrational acceleration required to attain this phase increases with bed thickness. This phenomenon is similar to the onset of bubbling \cite{pak1994}, where collection of voids form at the base and lift the material above. In cohesive powders,  at the onset of DF regime, we observed the widening of lower most crack (near the vibrated base) and lifting the powder column upward. Similar to upward traveling collection of bubbles in coarse grains, the waves due to successive crack widening seem to travel upwards as the vibrational strength is increased. Also, the onset of DF regime increases with the column thickness. 
These qualitative similarities suggests a possibility of defining a general governing mechanism within a broad range of granular materials.
\par Both Pak \textit{et al.} \cite{pak1994} and Eshuis \textit{et al.} \cite{eshuis2007} observed counter rotating convection rolls in their study of group B and group D powders, respectively. The former observed it at low vibrational strength and before bubbling, and the latter observed it at very high vibrational strengths. However, the bed height in both cases was different. In our case of fine group C powders, for flat vibrated bases, we only observed localized convection at the top surface at high frequencies as shown in Fig. \ref{fig:tvs}(f-g). In case of C1HS, the convection/deformation observed inside the container has random rotational direction. However, the upward moving powder slides past the wall. Thus, it doesn't form a heap at the top surface. We believe the convection failure in cohesive powder column is governed by some other mechanism. This is the topic of future investigation. 
\par In summary, we related phases observed in fine cohesive powders with more general granular material. Even though the phase behavior is different, the phase transitions occurs at similar $S$ and $\Gamma$ values. This comparison may help to find generalized mechanism to cause fracture/fluidization and phase transitions in cohesive powders.
%
%

\section{Conclusion}
\label{con}
\par We conclude that the group C powders under vibrations goes through four distinct phases: consolidated (CS), static fracture (SF), dynamic fracture (DF), and convective fracture (CF). 
We observed a brittle nature of vibrated fine cohesive powders, which is in contrast with the simple fluidized behavior of the vibrated coarse granular material. The brittle fracturing in consolidated cohesive powder is induced when vibrational strength exceed the thresholds. This boundary between the CS and SF/DF/CF phases, is controlled by the vibrational strength $S$. However, in DF phase, the effective decompaction wave propagation velocity is independent of $S$ and is governed by the gravity and cohesive strength.
\par We observed that the shape and size of container have minor effect on wave velocity. Although the shape dependence is not very significant, even such minor improvements might be helpful to design the energy-efficient system in industrial applications. Our investigation suggests that the hemispherical-based circular containers, carry out the most homogenized fluidization in group C powders. The fracturing mechanism is assisted by an unique phenomenon of powder slippage over the base and side-walls that promote convection in containers. We also report that the agglomerates formation and growth is possible when the powder is slide past against the curved walls of circular cross sections. In contrast, the flat side walls averse the agglomerate growth. However, flat walls tend to assist powder deposition. Thus, a role of the container's shape is not limited and can be explored to investigate various phenomenon and to gain industrial advantage. 
\par We conclude that fracturing in fine cohesive powders depends on $S$. However, the dependence on $S$ reduces in post-fracture regimes. This study can help to optimize geometrical parameters and the vibrational strength $S$ required for effective fluidization of fine cohesive group C powders with aeration, which otherwise are difficult to fluidize. 
%
%
%
%
\section{Acknowledgments}
\label{Ackn}
\par This work was supported by the JSPS KAKENHI, Grant No. 18H03679.
%
%
%
%
%
%
%
%

\bibliographystyle{elsarticle-num} 
\bibliography{bibPT}

\end{document}